\documentclass[11pt]{article}

\usepackage[preprint]{acl}
\usepackage{xurl}

\usepackage{tgtermes}
\usepackage[T1]{fontenc}
\usepackage[utf8]{inputenc}
\DeclareUnicodeCharacter{2212}{\textminus}
\usepackage{microtype}
\usepackage{inconsolata}
\usepackage{graphicx}
\usepackage{booktabs}
\usepackage{multirow}
\usepackage{amsmath}
\usepackage{xcolor}
\usepackage{colortbl}
\usepackage[most]{tcolorbox}
\newenvironment{promptbox}[2]{%
  \begin{tcolorbox}[
    enhanced jigsaw,
    breakable,
    colframe=#2!75!black,
    colback=#2!5!white,
    coltitle=white,
    fonttitle=\bfseries\footnotesize,
    boxrule=0.6pt,
    arc=2.5mm,
    boxsep=2pt,
    left=4pt, right=4pt, top=1pt, bottom=1pt,
    title=#1,
  ]%
  \scriptsize\raggedright}{%
  \end{tcolorbox}}

\title{Who Should Be on the Panel? Task-Conditioned Persona Selection for LLM Trading Agents}

\author{Boming Yang\textsuperscript{*}, Yuquan Wang\textsuperscript{*}, Jiaxian Guo, Yusuke Iwasawa, Yutaka Matsuo\\
  Graduate School of Engineering, The University of Tokyo\\
  \texttt{boming.yang@weblab.t.u-tokyo.ac.jp}}

\begin{document}
\maketitle
{\renewcommand{\thefootnote}{\fnsymbol{footnote}}\footnotetext[1]{Equal contribution.}}

\begin{abstract}
Multi-agent LLM systems are built by fixing a cast of roles and then designing how the cast coordinates. Which agents should make up the group for a given task is left to the designer, and the same cast is reused for every task. Financial trading makes the cost of that choice visible: stocks trade differently, and the trading styles suited to one need not suit another. We study task-conditioned persona selection, composing the group from the characteristics of the task it will face, and instantiate it with \textbf{FinPanel}, an LLM multi-agent trading system in which a recruiter selects a panel of named trader personas for each stock from a direction-blind statistical sketch of how the stock trades, each persona gives an independent daily call in its own style, and a manager commits the trade after reading the panel's arguments and their desk-kept, hierarchical track records. Because nothing is returned to the traders, the panel stays the panel that was selected, and it can trade a new stock from an empty record. On eight US stocks over the out-of-knowledge year 2025, against three prior LLM trading agents re-implemented in the same framework and against rule-based, predictive and reinforcement-learning baselines, the selected panel is ahead of a fixed panel of the same size and of the prior agents run as their papers prescribe, with a shallower drawdown than buy-and-hold on every stock; it keeps its performance with no warm-up and with no record at all, whereas the memory-augmented agents' returns swing with their warm-up; and neither more seats nor the manager's reading of the arguments substitutes for selecting the right ones. Who is in the group is a design decision to be made per task, before how the group coordinates.\footnote{Our code is available at\\ \url{https://github.com/tinyrolls/FinPanel}.}
\end{abstract}

\section{Introduction}

Autonomous agents built on large language models (LLMs) are increasingly organised as groups: several agents with distinct roles that debate, vote or hand work to one another \citep{mad,metagpt,agentverse}. The design of such a group has two parts, who is in it and how it coordinates, and research has concentrated on the second. A system fixes a cast, a set of analysts, critics or debaters chosen by the designer, and then studies how the cast should talk; which agents should make up the group for the task at hand is seldom a decision the system makes.

Financial trading is a setting in which that decision has a visible price. A trading decision has to be made every day from prices, technical indicators, news and filings, under uncertainty, and stocks trade differently, so the style that suits one need not suit another \citep{sharpe1992style}. The industry is organised around this fact: a multi-manager platform runs many portfolio managers who trade independently, each in its own style, and allocates capital among them on the record each has earned \citep{gs2024multimanager}. For a stock such as TSLA, a desk would seat a tail-risk trader and a trend follower rather than a value investor, and would judge each by what its calls earned.

Recent LLM agents for single-stock trading leave both parts of the design fixed. Memory-augmented single agents such as FinMem \citep{finmem} and FinAgent \citep{finagent} give one agent a character chosen by the designer and a self-reflective memory of its own trades; role-based multi-agent systems such as TradingAgents \citep{tradingagents} and FinCon \citep{fincon} model a trading firm with a fixed cast of analysts, bull and bear researchers, a trader and risk officers. In either case the same cast trades every stock, and no step of the system asks which traders suit this stock. What composition there is does not hold, either: following Reflexion \citep{reflexion}, what an agent learns is returned to the agent that learned it, so the agent is reshaped by its history on the stock, has to be warmed up on months or years of that history before its memory holds anything, and starts a new stock from nothing. Neither decision is made for the stock.

To address these limitations, we propose \textbf{FinPanel}, an LLM multi-agent trading system that composes a panel of traders for each stock the way a desk does. Our starting point is that the name of a well-known trader is a style the backbone already holds: asked to trade as a tail-risk investor or as a trend follower, the model behaves accordingly from the first day, without any history of the stock. FinPanel exploits this through four roles. A \emph{recruiter} reads a direction-blind statistical sketch of how the stock trades and seats $N$ named real traders suited to it. Each \emph{trader} reads the same daily information and gives an independent call with an argument, in its own style. A \emph{distiller} settles each call against the next day's market and maintains a hierarchical track record of each trader over month, quarter, half-year and year, which is kept for the manager and never returned to the trader, so that each trader keeps the style it was selected for. A \emph{manager} reads the arguments and the records and commits the portfolio to one action. Because a trader's call depends neither on the panel nor on any record, the panel stays the panel that was selected, can start on a new stock from an empty record, and can be taken apart and evaluated on the same trader calls.

We evaluate FinPanel on eight US stocks over the full year 2025 and its first half, a test period after the knowledge cutoff of the backbone we use, against FinMem, FinAgent and TradingAgents re-implemented in the same framework and against rule-based, predictive and reinforcement-learning baselines. Run as their papers prescribe, the three prior systems trail FinPanel, which returns 16.1\% on the mean of the eight stocks against 7.9\% for the best of them, and FinPanel has a shallower maximum drawdown than buy-and-hold on all eight stocks. The ablations, each paired on identical trader calls, then ask where the gain comes from. A fixed panel of the same six names loses most of it, so it is not from having traders; a panel selected from the ticker alone lands with the fixed panel, so it is from the sketch; eight seats do no better than six and the manager adds only a few points over the majority of the same calls, so neither more members nor a stronger aggregator substitutes for selecting the right ones; and the panel keeps its performance with no warm-up and with no record at all, so it is not from history on the stock, whereas the memory-augmented agents' returns swing with their warm-up. We call the gain the \emph{selection effect}.

Our contributions are summarised as follows:
\begin{itemize}
  \item We frame task-conditioned persona selection, choosing who is in a multi-agent group from a description of the task that contains no outcome, as a design decision in its own right, and instantiate it in financial trading with FinPanel.
  \item Through controlled ablations on identical trader calls, we show that selecting the traders for the stock accounts for the gain over a fixed panel, that neither more seats nor the manager's aggregation substitutes for it, and that the panel keeps its performance from an empty record on a new stock, or with no record at all.
  \item We provide a controlled evaluation on a public backtesting benchmark over an out-of-knowledge year, against three prior LLM trading agents re-implemented in our framework and run under both warm-up regimes; the framework, the re-implementations and all prompts are open-sourced.
\end{itemize}

\section{Related Work}

\subsection{LLM Agents for Financial Trading}
LLM-based trading agents make daily decisions on a single stock from prices, technical indicators, news and filings. One line of work equips a single agent with memory and reflection. FinMem \citep{finmem} layers a memory of market events and the agent's own reflections under a character whose risk preference is set by the designer; FinAgent \citep{finagent} adds tool use and a two-level reflection over past trades. These systems follow the Reflexion pattern \citep{reflexion}, in which what the agent learns is returned to the agent that learned it, so the memory has to be warmed up on the stock before it holds anything, and the reported results include the warm-up. A second line organises several agents in the roles of a trading firm. TradingAgents \citep{tradingagents} assigns analyst, bull and bear researcher, trader and risk-management roles and lets them debate to a decision; FinCon \citep{fincon} coordinates a manager and analysts with verbal reinforcement; QuantAgents \citep{quantagents} fixes a cast of specialised agents around a simulated desk. In these systems the views are assigned by role, a bull and a bear researcher answer different questions by construction, and the cast is designed once and reused for every stock. FinPanel differs from both lines in who trades: its traders are named real traders selected for the stock, every trader answers the same question in its own style, and the record of each trader is kept for the manager rather than returned to the trader. FINSABER \citep{finsaber}, a backtesting benchmark that re-evaluates LLM trading strategies under strict time ordering and trading costs, finds that their reported advantages shrink on longer windows and wider universes; we evaluate under its protocol (\S\ref{sec:setup}).

\subsection{Composition and Coordination in Multi-Agent LLM Systems}
A large body of work studies how a group of LLM agents should be composed and how it should coordinate. On composition, MetaGPT \citep{metagpt} encodes roles as standard operating procedures, AgentVerse \citep{agentverse} has an LLM write the expert descriptions for each task with the number of experts fixed by hand, DyLAN \citep{dylan} selects agents by an importance score measured inside the task, and AgentInit \citep{agentinit} initialises the group for diversity and expertise before it runs. On coordination, assigning a persona in the system prompt does not raise accuracy on tasks with a verifiable answer \citep{personas_not_helpful}, sampling more answers and voting improves with scale \citep{more_agents}, and controlled studies of multi-agent debate \citep{mad} find that most of its gain comes from majority aggregation and that debate can suppress a correct minority \citep{debate_or_vote}; recent work recovers the gain by starting the debate from a more diverse set of answers \citep{zhu2026demystifying}, and finds that agents which share information but decide independently adapt better in changing environments \citep{wu2025disagreement}, while samples from one model collapse toward a mode \citep{verbalized_sampling} and diversity helps where it changes what is considered \citep{wang2024spp}. That an equal-weight combination of forecasts is hard to beat is a long-standing result in forecasting \citep{clemen1989}. These studies ask what coordination buys once the group exists, on tasks that have a verifier, where a persona adds little because there is one answer to converge on. Trading has no verifier at decision time and no single right style. We therefore take the composition as the variable: we select named personas from a description of the task that contains no outcome information, hold the selection fixed, let each persona decide independently, and keep each agent's record away from it, so that a long history on the stock does not press the panel's styles into one.

\section{FinPanel}
\label{sec:system}

In a trading firm, a desk does not ask one trader to be right on every stock. A multi-manager platform runs many portfolio managers who trade independently, each in its own style, and allocates capital among them on the record each has earned \citep{gs2024multimanager}; a desk staffs each book the same way, a trend follower on one name and a tail-risk trader on another, and keeps a track record of each trader that the head of the desk reads before committing the book. The traders are not asked to become one another: a trend follower who has been wrong for a quarter is still a trend follower, and it is the desk, not the trader, that decides how much that quarter should weigh. FinPanel is built on this organisation. It separates two decisions that existing LLM trading agents fix by design, who trades the stock and who keeps the record of how the trades turned out, and gives each to a role of its own. The second decision is what keeps the panel a panel. A memory returned to the trader, as in the self-reflective agents, reshapes the trader with its own history on the stock; a record kept for the manager is designed to leave each trader judging in the style it was selected for, and to leave to the manager how much weight that judgement has earned. The traders stay traders; the desk does the learning.

\begin{figure*}[t]
  \centering
  \includegraphics[width=\textwidth]{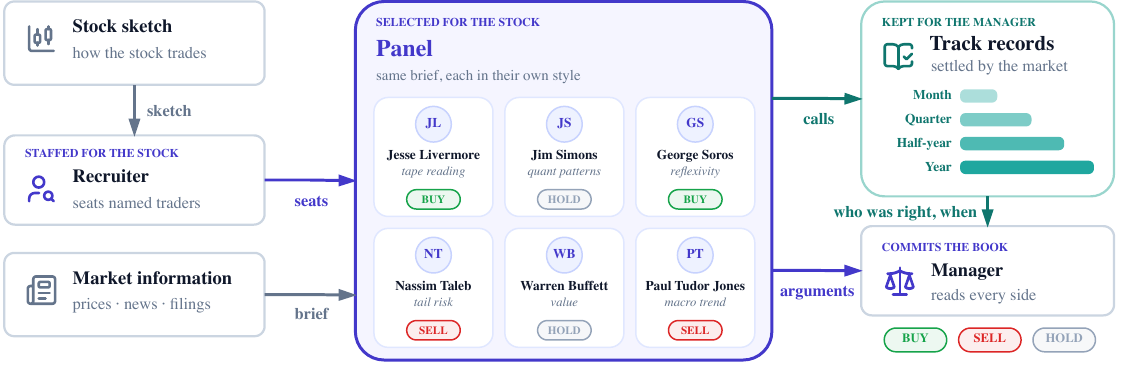}
  \caption{Overview of FinPanel. The recruiter reads a direction-blind sketch of the stock and seats $N$ named traders once per run; every day each trader reads the same brief and gives an independent call with an argument; the distiller settles each call against the market and maintains a hierarchical track record per trader; the manager reads the arguments and the records and commits the portfolio. The six names shown are the panel recruited for TSLA.}
  \label{fig:framework}
\end{figure*}

\subsection{Preliminaries}
\label{sec:problem}
We consider single-stock daily trading, the setting of the LLM trading agents we compare with \citep{finmem,finagent,tradingagents}. On each trading day $t$ the system reads what is known up to that day, prices, technical indicators, news and filings, and commits to one action $a_t\in\mathcal{A}=\{\text{BUY},\text{SELL},\text{HOLD}\}$: the portfolio is either fully in the stock or fully in cash, and HOLD keeps it where it is. Execution, trading costs and the starting capital follow the benchmark (Appendix~\ref{app:details}).

\subsection{Overview}
\label{sec:overview}
Figure~\ref{fig:framework} follows one stock through a run. Before the first trading day, the \emph{recruiter} reads a sketch $s$ of how the stock has traded and seats a panel $P=\{p_1,\dots,p_N\}$ of $N$ named real traders whose styles suit it; the panel then stands for the whole run. On each trading day the \emph{traders} read the same brief $b_t$, prices, indicators, news and filings cut at the decision day, and each trader $p_i$ argues the day in its own style and calls $a^i_t\in\mathcal{A}$ with an argument $e^i_t$. The \emph{manager} reads the $N$ arguments together with the track record $R^i_t$ the desk keeps on each trader and commits the portfolio to one action $a_t$. Once the next day's open is known, the \emph{distiller} settles every call against the realised move and updates each trader's record, which the manager reads the next day. Nothing flows back to the traders: a trader's call on day $t$ is made under the same prompt as on day one. This is why the panel can start on a new stock with an empty record, and why every other component can be removed and evaluated on the same trader calls (\S\ref{sec:experiments}).

\subsection{Roles}
\label{sec:roles}
\paragraph{Recruiter.}
The recruiter decides who trades the stock. Its input is a \emph{stock sketch} $s$: direction-blind statistics of how the stock itself trades, unchanged if every return is negated, so the sketch says how the stock moves but not which way it went. From $s$ the recruiter names $N$ real, well-known traders or investors suited to the stock and varied in how they trade. A trader is a name, and the name alone carries the style: the recruiter writes no profile, so that the persona is not locked to a description of our making, and what the backbone already knows of the name is what it trades with. The sketch and the panel recruited for TSLA are given in Appendix~\ref{app:roster}.

\paragraph{Traders.}
Each trader is one independent view of the day. All traders read the same brief, and each argues the day in its own style and calls BUY, SELL or HOLD. A trader sees neither the other traders nor the portfolio nor any record, so its call depends only on the brief and the name it trades under: a trader seated on two panels gives the same call on both, and a trader's first day on a stock is as informed as its hundredth.

\paragraph{Distiller.}
The distiller keeps the desk's record of each trader. Once the next day's market is known, it settles every trader's call of the previous day against the realised move and writes the outcome into that trader's \emph{hierarchical track record}: one bucket per month, quarter, half-year and year on the stock, each coarser bucket distilled from the finer ones into a characterisation of when the trader's calls tracked the move and when they missed it. Unlike the self-reflection memory of Reflexion and FinMem \citep{reflexion,finmem}, the record is written about a trader and read only by the manager; nothing is returned to the trader. The record may therefore start empty: with no warm-up, the first bucket is written after the first settled day, and nothing in the system requires a trading record on the stock.

\paragraph{Manager.}
The manager commits the portfolio. Each day it reads the $N$ arguments with their calls and the track record of each trader, and commits to one action $a_t$. It is asked to weigh the arguments against the records rather than to count the calls; how much this adds over the majority of the calls is measured in \S\ref{sec:panel}.

\section{Experiments}
\label{sec:experiments}
In this section we evaluate FinPanel to answer the following research questions. \textbf{RQ1}: How does a panel selected for the stock perform against prior LLM trading agents and classic baselines? \textbf{RQ2}: Does who is on the panel matter, and does it have to be selected for the stock? \textbf{RQ3}: Does the gain come from the selection or from history on the stock? \textbf{RQ4}: Can more seats or a stronger aggregator substitute for the selection?

\subsection{Experimental Setup}
\label{sec:setup}

\paragraph{Benchmark and data.}
We build on FINSABER \citep{finsaber}, a public backtesting benchmark for LLM trading strategies that enforces strict time ordering and charges trading costs on every order, and adopt its data, execution rules and cost model, so that every method runs on the same inputs. Its data run to the end of 2025, so we take the most recent full year, 2025, as the test period, after a two-year training and warm-up segment. We evaluate on eight US stocks: TSLA, NFLX, AMZN and MSFT, the stocks of the FinMem and FinAgent papers, and GS, WMT, BA and XOM, one from each of four further sectors. Every method trades the full year, and we report two windows: 2025-H1, the first half, which contains the April 2025 tariff shock and the recovery that followed, and 2025-FY, the full calendar year. Data sources, the stock selection rule and the window definitions are given in Appendix~\ref{app:details}.

\paragraph{Backbone.}
With 2025 as the test period, we use a backbone whose knowledge cutoff precedes it: gpt-oss-120b \citep{gptoss}, cutoff June 2024, run at temperature 0. Identical prompts return identical replies in over 99.8\% of calls across independent runs (Appendix~\ref{app:repro}), so every result in the paper is reproducible from its inputs. Every LLM method runs on this backbone.

\paragraph{Baselines.}
We compare with four groups of baselines, as in the single-stock agent papers; the first three are FINSABER's implementations under its training protocol. \emph{Market}: buy-and-hold. \emph{Rule-based}: SMA crossover, WMA crossover, Bollinger bands, trend following, ATR band and turn-of-month. \emph{Predictor}: ARIMA and XGBoost \citep{xgboost} as directional predictors. \emph{RL}: A2C, PPO, SAC and TD3 \citep{a2c,ppo,sac,td3}. \emph{LLM agents}: FinMem, FinAgent and TradingAgents, each re-implemented in our framework with the same inputs and cost model; FinMem and FinAgent are run with the two-year warm-up their papers prescribe and, for RQ3, with none. Unless stated otherwise, FinPanel seats $N{=}6$ traders and starts with the two-year record.

\paragraph{Evaluation metrics.}
We report cumulative return (CR, \%), the annualised Sharpe (SR) and Sortino (SoR) ratios, and maximum drawdown (MDD, \%); CR is the primary metric and MDD the secondary. Every result is reported per stock, with the mean over the eight stocks for summary.

\subsection{Overall Performance Comparison (RQ1)}
\label{sec:overall}

\begin{figure*}[t]
  \centering
  \includegraphics[width=\textwidth]{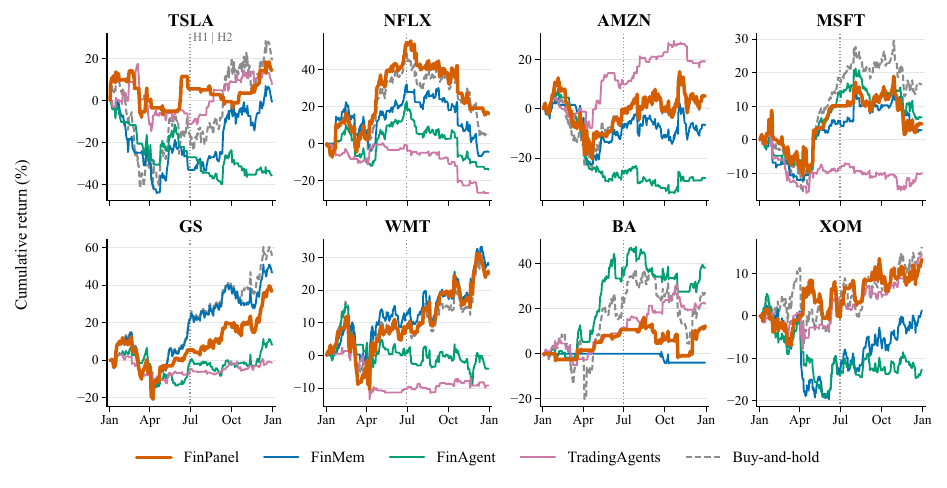}
  \caption{Cumulative return over 2025-FY per stock for FinPanel, the three LLM agents run as their papers prescribe, and buy-and-hold.}
  \label{fig:equity}
\end{figure*}

\begin{table*}[t]
  \centering\footnotesize\setlength{\tabcolsep}{2.55pt}
  \begin{tabular}{llrrrrrrrrrrrr}
\toprule
 & & \multicolumn{4}{c}{FinMem/FinAgent stocks} & \multicolumn{4}{c}{Added stocks} & \multicolumn{4}{c}{Mean over the eight stocks} \\
\cmidrule(lr){3-6}\cmidrule(lr){7-10}\cmidrule(lr){11-14}
Category & Method & TSLA & NFLX & AMZN & MSFT & GS & WMT & BA & XOM & CR\%$\uparrow$ & SR$\uparrow$ & SoR$\uparrow$ & MDD\%$\uparrow$ \\
\midrule
\multirow{1}{*}{Market} & Buy-and-hold & 19.2 & 5.0 & 4.5 & 16.7 & 55.8 & 25.4 & 27.4 & 16.2 & 21.3 & 0.68 & 0.96 & −28.3 \\
\midrule
\multirow{6}{*}{Rule-based} & SMA cross & 16.1 & 8.8 & −7.5 & 13.8 & 34.4 & −3.2 & 1.7 & 6.4 & 8.8 & 0.34 & 0.48 & −21.0 \\
 & WMA cross & 26.8 & −4.5 & −9.2 & 5.0 & 41.4 & 1.3 & 17.4 & −8.6 & 8.7 & 0.25 & 0.39 & −18.5 \\
 & Bollinger & 5.3 & 29.4 & 9.2 & 10.2 & 6.2 & 4.7 & 35.9 & 13.3 & 14.3 & 0.63 & 1.16 & −15.2 \\
 & Trend following & 6.2 & 6.1 & 12.4 & 16.0 & 47.9 & 10.1 & −9.7 & −28.3 & 7.6 & 0.16 & 0.38 & −15.6 \\
 & ATR band & 45.0 & 14.6 & 8.5 & 5.1 & 10.8 & 27.3 & 42.9 & 23.8 & 22.3 & 0.79 & 1.26 & −20.5 \\
 & Turn of month & −28.4 & −18.2 & −16.0 & −7.7 & −13.6 & 11.1 & −15.7 & −20.9 & −13.7 & −0.94 & −1.03 & −24.4 \\
\midrule
\multirow{2}{*}{Predictor} & ARIMA & 7.0 & −7.8 & −4.6 & 8.8 & 49.5 & 20.6 & 44.3 & −0.1 & 14.7 & 0.55 & 0.85 & −22.5 \\
 & XGBoost & −9.2 & 19.5 & 13.3 & 12.5 & −0.7 & 21.9 & −0.6 & 22.5 & 9.9 & 0.54 & 1.08 & −18.3 \\
\midrule
\multirow{4}{*}{RL} & A2C & 2.6 & −0.1 & −1.0 & 14.7 & 28.6 & 8.0 & 19.4 & 4.2 & 9.6 & 0.47 & 0.67 & −14.6 \\
 & PPO & 14.3 & 4.5 & 15.6 & 0.0 & 43.2 & 12.5 & 17.5 & 7.2 & 14.3 & 0.55 & 0.77 & −16.7 \\
 & SAC & 19.2 & 4.5 & 4.5 & 16.7 & 55.4 & 22.7 & 27.3 & 16.2 & 20.8 & 0.67 & 0.95 & −27.8 \\
 & TD3 & 19.2 & 4.5 & 4.5 & 16.7 & 55.4 & 22.7 & 27.3 & 16.2 & 20.8 & 0.67 & 0.95 & −27.8 \\
\midrule
\multirow{3}{*}{LLM agents} & FinMem & −0.6 & \underline{−4.6} & −6.4 & 2.9 & \textbf{46.6} & \textbf{27.7} & −3.9 & 1.3 & \underline{7.9} & \underline{0.15} & \underline{0.26} & −23.2 \\
 & FinAgent & −35.7 & −14.2 & −27.9 & \textbf{6.7} & 8.1 & −4.0 & \textbf{38.2} & −12.6 & −5.2 & −0.28 & −0.32 & −25.1 \\
 & TradingAgents & \underline{7.7} & −26.7 & \textbf{19.2} & −9.8 & −1.2 & −9.2 & \underline{22.2} & \textbf{14.6} & 2.1 & −0.05 & 0.02 & \textbf{−17.3} \\
\midrule
\rowcolor{gray!15}\multirow{1}{*}{Ours} & FinPanel & \textbf{14.5} & \textbf{16.1} & \underline{5.3} & \underline{4.9} & \underline{37.2} & \underline{25.0} & 12.5 & \underline{13.3} & \textbf{16.1} & \textbf{0.60} & \textbf{0.82} & \underline{−20.7} \\
\bottomrule
\end{tabular}

  \caption{Performance comparison on 2025-FY. Cumulative return (\%) per stock and the mean CR, Sharpe ratio, Sortino ratio and maximum drawdown over the eight stocks (MDD: higher is shallower). Every method is in its published configuration, and Table~\ref{tab:ablwarm} reports the LLM agents without their warm-up; all LLM methods run on gpt-oss-120b with the same inputs and costs. Among the LLM methods, the best per column is in bold and the second best underlined; the FinPanel row is shaded.}
  \label{tab:main}
\end{table*}

\begin{table*}[t]
  \centering\footnotesize\setlength{\tabcolsep}{2.55pt}
  \begin{tabular}{llrrrrrrrrrrrr}
\toprule
 & & \multicolumn{4}{c}{FinMem/FinAgent stocks} & \multicolumn{4}{c}{Added stocks} & \multicolumn{4}{c}{Mean over the eight stocks} \\
\cmidrule(lr){3-6}\cmidrule(lr){7-10}\cmidrule(lr){11-14}
Category & Method & TSLA & NFLX & AMZN & MSFT & GS & WMT & BA & XOM & CR\%$\uparrow$ & SR$\uparrow$ & SoR$\uparrow$ & MDD\%$\uparrow$ \\
\midrule
\multirow{1}{*}{Market} & Buy-and-hold & −15.2 & 48.5 & 0.4 & 18.3 & 20.2 & 8.5 & 25.1 & 3.1 & 13.6 & 0.85 & 1.20 & −26.9 \\
\midrule
\multirow{6}{*}{Rule-based} & SMA cross & −10.8 & 28.9 & −0.6 & 20.9 & 3.9 & −6.8 & −0.7 & −1.1 & 4.2 & 0.54 & 1.11 & −18.9 \\
 & WMA cross & 15.4 & 11.3 & 0.4 & 4.7 & 15.4 & −1.3 & 10.5 & −13.0 & 5.4 & 0.38 & 0.53 & −16.7 \\
 & Bollinger & −4.0 & 36.5 & 1.6 & 11.8 & 4.7 & 4.7 & 33.6 & 13.3 & 12.8 & 1.07 & 2.48 & −13.4 \\
 & Trend following & 3.3 & 22.5 & 14.1 & 19.2 & 29.3 & 8.4 & −4.0 & −16.6 & 9.5 & 0.90 & 1.58 & −11.1 \\
 & ATR band & 6.0 & 23.5 & −2.3 & 8.4 & 1.0 & 6.5 & 29.1 & 12.5 & 10.6 & 0.74 & 1.25 & −20.2 \\
 & Turn of month & −38.1 & −10.8 & −18.9 & −3.0 & −18.6 & −4.5 & −26.2 & −18.1 & −17.3 & −1.85 & −1.95 & −23.4 \\
\midrule
\multirow{2}{*}{Predictor} & ARIMA & −19.2 & 8.1 & −7.2 & −0.3 & 22.3 & 3.6 & 46.5 & −7.1 & 5.8 & 0.43 & 0.91 & −20.6 \\
 & XGBoost & −18.7 & 19.5 & −14.1 & 2.0 & −3.1 & 21.9 & 11.2 & 11.6 & 3.8 & 0.46 & 1.09 & −18.3 \\
\midrule
\multirow{4}{*}{RL} & A2C & −12.9 & 21.0 & −2.9 & 15.6 & 9.4 & −0.0 & 18.2 & −1.9 & 5.8 & 0.68 & 1.09 & −12.8 \\
 & PPO & −9.0 & 27.3 & 7.4 & 5.6 & 15.7 & 3.5 & 16.0 & 0.1 & 8.3 & 0.71 & 0.98 & −17.3 \\
 & SAC & −15.2 & 43.2 & 0.4 & 18.3 & 20.1 & 7.6 & 25.0 & 3.1 & 12.8 & 0.85 & 1.20 & −26.4 \\
 & TD3 & −15.2 & 43.2 & 0.4 & 18.3 & 20.1 & 7.6 & 25.0 & 3.1 & 12.8 & 0.85 & 1.20 & −26.4 \\
\midrule
\multirow{3}{*}{LLM agents} & FinMem & −33.1 & \underline{30.6} & −6.1 & 5.5 & \textbf{19.3} & \textbf{12.1} & 0.0 & −12.7 & \underline{1.9} & \underline{0.15} & \underline{0.22} & −21.6 \\
 & FinAgent & −27.0 & 21.2 & −24.9 & \textbf{13.9} & −2.8 & 0.4 & \textbf{38.3} & −13.8 & 0.7 & −0.02 & 0.13 & −21.3 \\
 & TradingAgents & \underline{−11.1} & −0.8 & \textbf{11.8} & −8.5 & −7.8 & −10.2 & \underline{11.7} & \textbf{4.1} & −1.3 & −0.22 & −0.23 & \textbf{−15.3} \\
\midrule
\rowcolor{gray!15}\multirow{1}{*}{Ours} & FinPanel & \textbf{5.5} & \textbf{52.8} & \underline{−0.1} & \underline{10.2} & \underline{5.4} & \underline{8.0} & 10.8 & \underline{1.1} & \textbf{11.7} & \textbf{0.86} & \textbf{1.34} & \underline{−18.2} \\
\bottomrule
\end{tabular}

  \caption{Performance comparison on 2025-H1 (the April 2025 tariff shock and the recovery that followed); columns as in Table~\ref{tab:main}, with the Sharpe and Sortino ratios annualised from the half-year.}
  \label{tab:mainh1}
\end{table*}

Figure~\ref{fig:equity} shows the cumulative return paths of FinPanel, the three prior LLM agents and buy-and-hold over 2025-FY, and Tables~\ref{tab:main} and~\ref{tab:mainh1} compare FinPanel with all baselines on 2025-FY and on 2025-H1. We have the following observations. 1) FinPanel outperforms the three prior LLM agents. It returns 16.1\% on the mean of the eight stocks over 2025-FY against 7.9\% for the best of them run as their papers prescribe, and is ahead of each of them on most stocks and on the mean Sharpe and Sortino ratios; the picture on 2025-H1 is the same. On the four stocks the prior agents were built on, all three end 2025-FY with a loss on the mean of these four, while FinPanel returns 10.2\%, about a point behind buy-and-hold at a shallower drawdown on each, and in the stress window it is ahead of buy-and-hold on their mean. 2) Against buy-and-hold, FinPanel's advantage is the drawdown, not the return. Its drawdown is shallower on every one of the eight stocks, on both windows (Appendix~\ref{app:mdd}), as the April paths in Figure~\ref{fig:equity} show. In return, no LLM method, FinPanel included, beats buy-and-hold over 2025-FY, and two technical rules do better than FinPanel on risk-adjusted return, in line with the benchmark's finding that simple rules are hard for LLM strategies to beat \citep{finsaber}: in a year in which all eight stocks rose, every day spent in cash gives up part of the rise, and buy-and-hold gives up none. 3) The advantage over the prior agents is built in the stress window. The gap to the best of them is wider on 2025-H1 than on 2025-FY, and the TSLA panel of Figure~\ref{fig:equity} shows how: the panel selected for TSLA, a tail-risk trader among its seats, steps aside in the February--April decline and re-enters in the recovery, while FinMem and FinAgent ride the decline.

\subsection{Effect of Persona Selection (RQ2)}
\label{sec:selection}

\begin{table}[t]
  \centering\footnotesize\setlength{\tabcolsep}{2.5pt}
  \begin{tabular}{lrrrrrr}
\toprule
 & \multicolumn{3}{c}{2025-H1} & \multicolumn{3}{c}{2025-FY} \\
\cmidrule(lr){2-4}\cmidrule(lr){5-7}
Panel & CR$\uparrow$ & SR$\uparrow$ & MDD$\uparrow$ & CR$\uparrow$ & SR$\uparrow$ & MDD$\uparrow$ \\
\midrule
\rowcolor{gray!15}FinPanel & \textbf{11.7} & \textbf{0.86} & \textbf{−18.2} & \textbf{16.1} & \textbf{0.60} & \textbf{−20.7} \\
w/o selection & 2.6 & 0.13 & −22.9 & 7.2 & 0.24 & −25.1 \\
w/o sketch & 4.7 & 0.24 & −21.5 & 7.9 & 0.28 & −26.1 \\
\bottomrule
\end{tabular}

  \caption{Effect of persona selection: mean cumulative return (\%), Sharpe ratio and maximum drawdown (\%) over the eight stocks on 2025-H1 and 2025-FY for the recruited panel (FinPanel), the fixed panel seated on every stock (w/o selection), and the panel recruited from the ticker alone (w/o sketch). Best per column in bold; per-stock values are in Appendix~\ref{app:selcells}.}
  \label{tab:ablsel}
\end{table}

Table~\ref{tab:ablsel} compares the recruited panel with a fixed panel seated on every stock, the six names the recruiter seated most often across its recruited panels, and with the panel recruited from the ticker alone; a trader seated in more than one of these panels gives the same calls in each (\S\ref{sec:roles}), so the comparison is paired on the shared calls. We have the following observations. 1) The selection effect accounts for the gain. With the same six seats filled by the fixed panel, the mean return falls by about nine points on both windows and the drawdown deepens, and the recruited panel is ahead on most stocks. The two panels share most of their seats (Appendix~\ref{app:roster}), so the effect is carried by the one or two seats filled differently for a stock. 2) The gain comes from the sketch, not from the name of the stock. Given the ticker alone, the recruiter seats a panel that lands with the fixed panel on both windows, and the recruited panel is ahead of it on nearly every stock. 3) The gap is built in the stress window and carried through the calmer second half: the largest per-stock gaps are TSLA and AMZN on 2025-H1, and the gap at the end of 2025-FY is the 2025-H1 gap carried through.

\subsection{Effect of the Track Record and Warm-up (RQ3)}
\label{sec:warmup}

\begin{table}[t]
  \centering\footnotesize\setlength{\tabcolsep}{1.6pt}
  \begin{tabular}{lrrrrrr}
\toprule
 & \multicolumn{3}{c}{2025-H1} & \multicolumn{3}{c}{2025-FY} \\
\cmidrule(lr){2-4}\cmidrule(lr){5-7}
Method & CR$\uparrow$ & SR$\uparrow$ & MDD$\uparrow$ & CR$\uparrow$ & SR$\uparrow$ & MDD$\uparrow$ \\
\midrule
\rowcolor{gray!15}FinPanel, 2-y & \underline{11.7} & \underline{0.86} & −18.2 & 16.1 & \underline{0.60} & −20.7 \\
\rowcolor{gray!15}FinPanel, zero & 11.5 & \underline{0.86} & −17.2 & 15.6 & \textbf{0.61} & \underline{−19.5} \\
\rowcolor{gray!15}FinPanel, no record & \textbf{13.1} & \textbf{0.87} & \underline{−16.3} & \underline{16.2} & 0.55 & −20.1 \\
\midrule
FinMem, 2-y & 1.9 & 0.15 & −21.6 & 7.9 & 0.15 & −23.2 \\
FinMem, zero & 10.5 & 0.70 & −23.8 & \textbf{18.6} & \underline{0.60} & −25.3 \\
\midrule
FinAgent, 2-y & 0.7 & −0.02 & −21.3 & −5.2 & −0.28 & −25.1 \\
FinAgent, zero & −0.2 & −0.06 & −20.7 & −0.9 & −0.03 & −25.5 \\
\midrule
TradingAgents & −1.3 & −0.22 & \textbf{−15.3} & 2.1 & −0.05 & \textbf{−17.3} \\
\bottomrule
\end{tabular}

  \caption{Effect of the warm-up: mean cumulative return (\%), Sharpe ratio and maximum drawdown (\%) over the eight stocks on 2025-H1 and 2025-FY for each LLM method with the two-year warm-up its paper prescribes (2-y) and with none (zero), and for FinPanel also with the record removed (no record). Best per column in bold and second best underlined; the FinPanel rows are shaded; TradingAgents is memoryless. Per-stock values are in Appendix~\ref{app:warmcells}.}
  \label{tab:ablwarm}
\end{table}

Table~\ref{tab:ablwarm} reports each LLM method with its two-year warm-up and with none, and FinPanel also with no record at all, the manager reading the arguments alone. We have the following observations. 1) The gain is in the selection, not in the history on the stock: the panel keeps its performance without a warm-up, and without the record. The same six traders cast the same calls in every row and only what the manager reads differs; from the two-year record to an empty one the mean return moves by less than half a point, and with no record at all it is no lower. 2) The prior agents depend on their warm-up, and not in the way their papers intend. Without it FinMem's mean return more than doubles on both windows, and FinAgent moves the same way on 2025-FY. We read this as the memory written on 2023--2024 being a liability rather than an asset in a year whose first half carries the tariff shock. 3) Without its warm-up FinMem draws level with the panel in return, by a different route: it holds the stock on far more days and at a deeper drawdown (Appendix~\ref{app:mdd}), so in a rising year it takes more of buy-and-hold's return and more of its drawdown; FinAgent and TradingAgents remain behind the panel in both regimes.

\subsection{Effect of Panel Size and Aggregation (RQ4)}
\label{sec:panel}

\begin{figure}[t]
  \centering
  \includegraphics[width=\columnwidth]{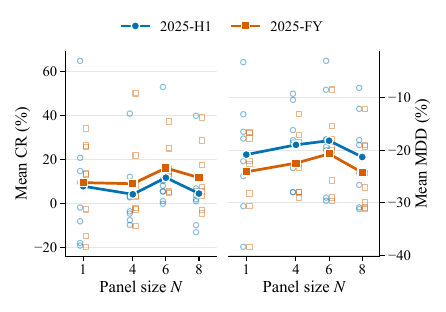}
  \caption{Mean cumulative return and maximum drawdown against panel size $N\in\{1,4,6,8\}$ on 2025-H1 and 2025-FY, with recruited panels and the two-year record; the light markers are the eight per-stock values. Per-stock values are in Appendix~\ref{app:size}.}
  \label{fig:size}
\end{figure}

Figure~\ref{fig:size} varies the panel size, $N\in\{1,4,6,8\}$, with everything else fixed, the recruiter seating one trader at $N{=}1$. We have the following observations. 1) More seats do not substitute for the right ones: six seats are ahead of four and of eight on both windows, with a shallower drawdown than eight on every stock. 2) A single selected trader is a bet on its style. At $N{=}1$ the recruiter seats its single best fit for each stock; it trails the panel by several points on the mean, has the widest per-stock spread in Figure~\ref{fig:size} on 2025-H1, and ends 2025-FY with a loss on three stocks where the panel has none. The trader that would have been best differs across stocks and windows (Appendix~\ref{app:solo}), so it could not have been chosen in advance; the panel is a way not to have to. 3) The manager mostly counts, and earns a little by not always counting. Reconstructing on every day the majority vote of the same calls, the manager's action equals it on most days, and the manager's book is a few points ahead of the vote's on both windows (Appendix~\ref{app:solo}), a difference carried by TSLA and BA and far smaller than the selection effect, so a stronger aggregator does not substitute for the selection either.

\section{Conclusion}
\label{sec:conclusion}
Who is on the panel decides the panel's return; how long it has been there does not. These are the findings of this paper, obtained with FinPanel, which selects a panel of named trader personas for each stock from a direction-blind sketch and has a manager read their hierarchical track records. On eight US stocks over the out-of-knowledge year 2025, the selected panel returns 16.1\% on the mean against 7.9\% for the best of three prior LLM trading agents run as their papers prescribe, with a shallower drawdown than buy-and-hold on every stock; a fixed panel of the same size loses most of the gain, more seats and a stronger aggregator add little, and the panel keeps its performance without a warm-up or any record, whereas the memory-augmented agents' returns swing with theirs. For the design of multi-agent LLM systems, the implication is that who is in the group is a decision to be made per task, before how the group coordinates, and that for a task without a verifier the members can be selected from a description of the task alone. The panel's advantage lies in the floor rather than the ceiling: in a rising year no LLM method, ours included, beats buy-and-hold in return.

\section*{Limitations}
\label{sec:limitations}
Our evaluation covers eight US stocks over one calendar year, 2025, which was a rising year for all eight with one stress episode in its first half; results are reported per stock and as means over the eight stocks. All LLM methods run on a single backbone, gpt-oss-120b. The three prior agents are re-implemented on our backbone, inputs and position sizing (Appendix~\ref{app:details}), so their absolute returns are not those of their papers. Trading is simulated: single-stock, long-only and binary, under the benchmark's execution and cost model, with the news and filings it supplies. The composition question is studied in one task, single-stock trading. Further stocks, years, backbones, tasks and a second stress window are left to future work.

\section*{Ethics Statement}
The panel members are the names of real, well-known traders and investors, used only as labels for the trading styles the backbone associates with them; the trader prompt addresses the model as the named trader, which is how a style is invoked, and every call, argument and track record attributed to a name in this paper and its artefacts is generated by the model and is not a statement, view or performance of the person named. All data are the public prices, news and filings of the benchmark, and no personal data beyond the public names is used. The system trades a simulated portfolio on historical data; nothing in this paper is investment advice.

\bibliography{refs}

\clearpage
\appendix
\section{Experimental Details}
\label{app:details}
\paragraph{Data and windows.} Prices are FINSABER's daily bars; news are the top five items per trading day by embedding relevance over a seven-day window; filings are summaries of the latest 10-K/10-Q; all are cut at the decision day and shared by every method. The training and warm-up segment runs from 2023-01-01 to 2024-12-31 and the test period, 2025-FY, from 2025-01-02 to 2025-12-30 (249 scored days); each method holds one portfolio across the whole period, starting with \$100k, with orders filled at the next open under the benchmark's commission and slippage. 2025-H1 is the first 121 scored days of the same run, reported as the short window.

\paragraph{Selection of the four added stocks.} The four stocks outside the FinMem/FinAgent set were chosen by a rule on the training segment: for each GICS sector not already represented by the four, take the benchmark stock with the most news items, then keep the four sectors with the highest such counts, which gives GS, WMT, BA and XOM.

\paragraph{Baselines.} The six rules, the two predictors and the four RL agents are FINSABER's implementations on the same data and cost model; the RL agents train on the two-year segment before the test period under its protocol, and SAC and TD3 both converge to an always-long policy on every stock, so their rows coincide and track buy-and-hold.

\paragraph{Re-implemented agents.} FinMem, FinAgent and TradingAgents are ported into our framework with their public prompts and memory designs, on the same backbone and data as FinPanel. Common to the three: gpt-oss-120b in place of the GPT-4-class backbones of their papers, the shared top-five news and filing summaries in place of their own feeds, and our single-stock, long-only, binary position. FinMem keeps its four-layer memory, retrieval score, self-adaptive character, warm-up pass and prompts; the per-article sentiment line of its prompt is omitted, as the benchmark's news carries no sentiment scores. FinAgent keeps its five-stage pipeline, three memory stores and the benchmark's verbatim prompt assembly; its price and trading charts are rendered as text tables rather than images, the tool-augmentation stage is omitted, and the warm-up runs the plain pipeline for two years rather than three with teacher-forced reflections. TradingAgents keeps its analyst, researcher, trader, risk and manager roles with their debate lengths, but not its reflection memory, so it is memoryless and its two warm-up regimes coincide; its analysts read the frozen inputs rather than live tools, and one model replaces its deep- and quick-thinking pair.

\section{Reproducibility}
\label{app:repro}
All LLM calls run at temperature zero, and the backbone is served locally under deterministic inference, so a reply does not depend on the other requests in its batch. Across independent runs weeks apart, identical prompts returned identical replies in over 99.8\% of calls, so a run is reproducible from its inputs; re-run on the same sketch, the recruiter returns the same panel. Because a trader's daily prompt carries no panel- or record-dependent input, a trader's calls are identical in every configuration that seats it, so every ablation in the paper is paired on the same trader calls. FinMem and FinAgent retrieve their memories with text-embedding-ada-002; three independent runs of FinMem's zero-warm-up year reproduce its mean CR within one point. The panel retrieves nothing by embedding. The code, the prompts, the re-implemented baselines and the run outputs behind every table and figure are released.

\section{Per-Stock Results}
\label{app:perticker}
The tables of this section give, per stock and in the columns of Table~\ref{tab:main}, the values behind the ablation tables and the figure of \S\ref{sec:experiments}.

\subsection{Persona Selection}
\label{app:selcells}
Table~\ref{tab:selcells} expands Table~\ref{tab:ablsel}: the recruited panel, the fixed panel and the panel recruited from the ticker alone.

\begin{table*}[t]
  \centering\footnotesize\setlength{\tabcolsep}{2.7pt}
  \begin{tabular}{llrrrrrrrrrrrr}
\toprule
 & & \multicolumn{4}{c}{FinMem/FinAgent stocks} & \multicolumn{4}{c}{Added stocks} & \multicolumn{4}{c}{Mean over the eight stocks} \\
\cmidrule(lr){3-6}\cmidrule(lr){7-10}\cmidrule(lr){11-14}
Window & Method & TSLA & NFLX & AMZN & MSFT & GS & WMT & BA & XOM & CR\%$\uparrow$ & SR$\uparrow$ & SoR$\uparrow$ & MDD\%$\uparrow$ \\
\midrule
\rowcolor{gray!15} & FinPanel & 5.5 & 52.8 & −0.1 & 10.2 & 5.4 & 8.0 & 10.8 & 1.1 & 11.7 & 0.86 & 1.34 & −18.2 \\
 & w/o selection & −16.9 & 39.7 & −15.4 & 7.5 & 8.0 & −4.9 & 1.1 & 1.3 & 2.6 & 0.13 & 0.16 & −22.9 \\
\multirow{-3}{*}{2025-H1} & w/o sketch & −9.4 & 52.7 & −10.4 & 3.3 & −0.6 & 9.5 & 0.3 & −7.4 & 4.7 & 0.24 & 0.44 & −21.5 \\
\midrule
\rowcolor{gray!15} & FinPanel & 14.5 & 16.1 & 5.3 & 4.9 & 37.2 & 25.0 & 12.5 & 13.3 & 16.1 & 0.60 & 0.82 & −20.7 \\
 & w/o selection & −17.0 & 8.5 & −10.4 & 4.3 & 44.9 & 10.3 & −1.9 & 18.7 & 7.2 & 0.24 & 0.29 & −25.1 \\
\multirow{-3}{*}{2025-FY} & w/o sketch & −6.6 & 10.8 & −6.7 & 0.1 & 39.1 & 24.9 & −4.7 & 6.3 & 7.9 & 0.28 & 0.38 & −26.1 \\
\bottomrule
\end{tabular}

  \caption{Effect of persona selection per stock: cumulative return (\%) per stock and the mean CR, Sharpe ratio, Sortino ratio and maximum drawdown over the eight stocks, on 2025-H1 and 2025-FY, for the rows of Table~\ref{tab:ablsel}.}
  \label{tab:selcells}
\end{table*}

\subsection{Warm-up}
\label{app:warmcells}
Table~\ref{tab:warmcells} expands Table~\ref{tab:ablwarm}: each LLM method with its two-year warm-up and with none.

\begin{table*}[t]
  \centering\footnotesize\setlength{\tabcolsep}{2.7pt}
  \begin{tabular}{llrrrrrrrrrrrr}
\toprule
 & & \multicolumn{4}{c}{FinMem/FinAgent stocks} & \multicolumn{4}{c}{Added stocks} & \multicolumn{4}{c}{Mean over the eight stocks} \\
\cmidrule(lr){3-6}\cmidrule(lr){7-10}\cmidrule(lr){11-14}
Window & Method & TSLA & NFLX & AMZN & MSFT & GS & WMT & BA & XOM & CR\%$\uparrow$ & SR$\uparrow$ & SoR$\uparrow$ & MDD\%$\uparrow$ \\
\midrule
\rowcolor{gray!15} & FinPanel, 2-y & 5.5 & 52.8 & −0.1 & 10.2 & 5.4 & 8.0 & 10.8 & 1.1 & 11.7 & 0.86 & 1.34 & −18.2 \\
\rowcolor{gray!15} & FinPanel, zero & 18.6 & 48.0 & −4.9 & 8.5 & 3.7 & 8.5 & 7.2 & 2.3 & 11.5 & 0.86 & 1.50 & −17.2 \\
\rowcolor{gray!15} & FinPanel, no record & 9.3 & 54.0 & −2.3 & 14.9 & 11.6 & 6.7 & 3.8 & 6.5 & 13.1 & 0.87 & 1.30 & −16.3 \\
 & FinMem, 2-y & −33.1 & 30.6 & −6.1 & 5.5 & 19.3 & 12.1 & 0.0 & −12.7 & 1.9 & 0.15 & 0.22 & −21.6 \\
 & FinMem, zero & −9.8 & 48.3 & −0.1 & 7.1 & 19.3 & 7.7 & 12.9 & −1.5 & 10.5 & 0.70 & 0.94 & −23.8 \\
 & FinAgent, 2-y & −27.0 & 21.2 & −24.9 & 13.9 & −2.8 & 0.4 & 38.3 & −13.8 & 0.7 & −0.02 & 0.13 & −21.3 \\
 & FinAgent, zero & −38.1 & 22.3 & −17.6 & 7.6 & −0.9 & −1.0 & 26.3 & −0.2 & −0.2 & −0.06 & −0.02 & −20.7 \\
\multirow{-8}{*}{2025-H1} & TradingAgents & −11.1 & −0.8 & 11.8 & −8.5 & −7.8 & −10.2 & 11.7 & 4.1 & −1.3 & −0.22 & −0.23 & −15.3 \\
\midrule
\rowcolor{gray!15} & FinPanel, 2-y & 14.5 & 16.1 & 5.3 & 4.9 & 37.2 & 25.0 & 12.5 & 13.3 & 16.1 & 0.60 & 0.82 & −20.7 \\
\rowcolor{gray!15} & FinPanel, zero & 15.3 & 16.6 & 6.9 & 7.1 & 33.4 & 20.9 & 4.0 & 20.9 & 15.6 & 0.61 & 0.86 & −19.5 \\
\rowcolor{gray!15} & FinPanel, no record & 10.1 & 16.8 & 3.1 & 17.2 & 50.0 & 19.1 & −8.2 & 21.5 & 16.2 & 0.55 & 0.79 & −20.1 \\
 & FinMem, 2-y & −0.6 & −4.6 & −6.4 & 2.9 & 46.6 & 27.7 & −3.9 & 1.3 & 7.9 & 0.15 & 0.26 & −23.2 \\
 & FinMem, zero & 26.6 & 5.3 & 2.5 & 5.9 & 54.4 & 26.9 & 15.0 & 12.0 & 18.6 & 0.60 & 0.81 & −25.3 \\
 & FinAgent, 2-y & −35.7 & −14.2 & −27.9 & 6.7 & 8.1 & −4.0 & 38.2 & −12.6 & −5.2 & −0.28 & −0.32 & −25.1 \\
 & FinAgent, zero & −39.4 & −17.4 & −13.9 & 5.0 & 14.7 & 12.0 & 22.1 & 10.1 & −0.9 & −0.03 & −0.03 & −25.5 \\
\multirow{-8}{*}{2025-FY} & TradingAgents & 7.7 & −26.7 & 19.2 & −9.8 & −1.2 & −9.2 & 22.2 & 14.6 & 2.1 & −0.05 & 0.02 & −17.3 \\
\bottomrule
\end{tabular}

  \caption{Effect of the warm-up per stock: columns as in Table~\ref{tab:selcells}, for the rows of Table~\ref{tab:ablwarm}.}
  \label{tab:warmcells}
\end{table*}

\subsection{Panel Size}
\label{app:size}
Table~\ref{tab:size} gives the values behind Figure~\ref{fig:size}, at $N\in\{1,4,6,8\}$ with recruited panels and the two-year record.

\begin{table*}[t]
  \centering\footnotesize\setlength{\tabcolsep}{2.7pt}
  \begin{tabular}{llrrrrrrrrrrrr}
\toprule
 & & \multicolumn{4}{c}{FinMem/FinAgent stocks} & \multicolumn{4}{c}{Added stocks} & \multicolumn{4}{c}{Mean over the eight stocks} \\
\cmidrule(lr){3-6}\cmidrule(lr){7-10}\cmidrule(lr){11-14}
Window & Method & TSLA & NFLX & AMZN & MSFT & GS & WMT & BA & XOM & CR\%$\uparrow$ & SR$\uparrow$ & SoR$\uparrow$ & MDD\%$\uparrow$ \\
\midrule
 & $N{=}1$ & −19.1 & 64.7 & −18.0 & 14.7 & −1.8 & 9.5 & 20.7 & −8.1 & 7.8 & 0.48 & 1.12 & −20.8 \\
 & $N{=}4$ & −7.6 & 40.7 & −3.6 & 2.8 & 12.1 & 3.2 & −4.6 & −9.7 & 4.2 & 0.12 & 0.22 & −19.0 \\
\rowcolor{gray!15} & $N{=}6$ & 5.5 & 52.8 & −0.1 & 10.2 & 5.4 & 8.0 & 10.8 & 1.1 & 11.7 & 0.86 & 1.34 & −18.2 \\
\multirow{-4}{*}{2025-H1} & $N{=}8$ & −13.0 & 39.8 & −9.8 & 5.7 & 1.0 & 1.9 & 3.7 & 6.8 & 4.5 & 0.29 & 0.41 & −21.3 \\
\midrule
 & $N{=}1$ & −19.7 & 33.9 & −14.7 & 13.0 & 26.4 & 25.7 & 13.6 & −2.6 & 9.5 & 0.29 & 0.42 & −24.1 \\
 & $N{=}4$ & −2.8 & 3.4 & −2.1 & 3.2 & 50.0 & 21.9 & −10.2 & 8.2 & 9.0 & 0.26 & 0.36 & −22.5 \\
\rowcolor{gray!15} & $N{=}6$ & 14.5 & 16.1 & 5.3 & 4.9 & 37.2 & 25.0 & 12.5 & 13.3 & 16.1 & 0.60 & 0.82 & −20.7 \\
\multirow{-4}{*}{2025-FY} & $N{=}8$ & 3.4 & 5.4 & −2.9 & 7.4 & 39.0 & 17.5 & −4.6 & 28.5 & 11.7 & 0.41 & 0.56 & −24.2 \\
\bottomrule
\end{tabular}

  \caption{Panel size per stock: columns as in Table~\ref{tab:selcells}, at $N\in\{1,4,6,8\}$ with recruited panels and the two-year record, behind Figure~\ref{fig:size}.}
  \label{tab:size}
\end{table*}

\subsection{Sharpe Ratio, Drawdown and Exposure}
\label{app:mdd}
Tables~\ref{tab:sharpe}, \ref{tab:mdd} and~\ref{tab:exposure} give, per stock, the Sharpe ratio, the maximum drawdown and the exposure, the share of scored days on which the book holds the stock, behind the risk statements of \S\ref{sec:overall} and \S\ref{sec:warmup}: buy-and-hold, the three prior agents under both warm-up regimes, and FinPanel with the two-year record, with an empty one and with none.

\begin{table*}[t]
  \centering\footnotesize\setlength{\tabcolsep}{4pt}
  \begin{tabular}{llrrrrrrrrr}
\toprule
 & & \multicolumn{4}{c}{FinMem/FinAgent stocks} & \multicolumn{4}{c}{Added stocks} & \\
\cmidrule(lr){3-6}\cmidrule(lr){7-10}
Window & Method & TSLA & NFLX & AMZN & MSFT & GS & WMT & BA & XOM & Mean \\
\midrule
 & Buy-and-hold & −0.11 & 2.33 & 0.13 & 1.22 & 1.12 & 0.63 & 1.21 & 0.27 & 0.85 \\
 & FinMem, 2-y & −1.64 & 1.79 & −0.44 & 0.53 & 1.08 & 0.97 & 0.00 & −1.12 & 0.15 \\
 & FinMem, zero & 0.03 & 2.59 & −0.02 & 0.59 & 1.08 & 0.68 & 0.73 & −0.09 & 0.70 \\
 & FinAgent, 2-y & −1.42 & 1.64 & −2.77 & 1.32 & −0.32 & −0.00 & 2.99 & −1.64 & −0.02 \\
 & FinAgent, zero & −2.99 & 1.68 & −1.60 & 0.70 & −0.06 & −0.19 & 2.07 & −0.05 & −0.06 \\
 & TradingAgents & −0.69 & −0.17 & 0.91 & −1.17 & −0.96 & −1.47 & 1.40 & 0.40 & −0.22 \\
\rowcolor{gray!15} & FinPanel, 2-y & 0.41 & 2.72 & 0.09 & 0.77 & 0.40 & 0.61 & 1.85 & 0.07 & 0.86 \\
\rowcolor{gray!15} & FinPanel, zero & 1.71 & 2.41 & −0.31 & 0.65 & 0.31 & 0.64 & 1.25 & 0.19 & 0.86 \\
\rowcolor{gray!15}\multirow{-9}{*}{2025-H1} & FinPanel, no record & 0.80 & 2.68 & −0.05 & 1.10 & 0.75 & 0.53 & 0.60 & 0.56 & 0.87 \\
\midrule
 & Buy-and-hold & 0.55 & 0.23 & 0.22 & 0.64 & 1.49 & 0.92 & 0.76 & 0.64 & 0.68 \\
 & FinMem, 2-y & 0.13 & −0.11 & −0.19 & 0.08 & 1.31 & 1.07 & −1.12 & 0.03 & 0.15 \\
 & FinMem, zero & 0.64 & 0.23 & 0.10 & 0.24 & 1.47 & 1.10 & 0.48 & 0.52 & 0.60 \\
 & FinAgent, 2-y & −1.14 & −0.65 & −1.39 & 0.29 & 0.35 & −0.28 & 1.52 & −0.91 & −0.28 \\
 & FinAgent, zero & −1.49 & −0.80 & −0.66 & 0.20 & 0.61 & 0.54 & 0.89 & 0.45 & −0.03 \\
 & TradingAgents & 0.31 & −1.70 & 0.84 & −0.86 & −0.19 & −0.91 & 1.24 & 0.86 & −0.05 \\
\rowcolor{gray!15} & FinPanel, 2-y & 0.52 & 0.56 & 0.23 & 0.19 & 1.15 & 0.95 & 0.58 & 0.61 & 0.60 \\
\rowcolor{gray!15} & FinPanel, zero & 0.79 & 0.56 & 0.28 & 0.28 & 1.08 & 0.80 & 0.14 & 0.91 & 0.61 \\
\rowcolor{gray!15}\multirow{-9}{*}{2025-FY} & FinPanel, no record & 0.45 & 0.57 & 0.16 & 0.68 & 1.52 & 0.75 & −0.67 & 0.96 & 0.55 \\
\bottomrule
\end{tabular}

  \caption{Sharpe ratio per stock and its mean over the eight stocks, on 2025-H1 and 2025-FY.}
  \label{tab:sharpe}
\end{table*}

\begin{table*}[t]
  \centering\footnotesize\setlength{\tabcolsep}{4pt}
  \begin{tabular}{llrrrrrrrrr}
\toprule
 & & \multicolumn{4}{c}{FinMem/FinAgent stocks} & \multicolumn{4}{c}{Added stocks} & \\
\cmidrule(lr){3-6}\cmidrule(lr){7-10}
Window & Method & TSLA & NFLX & AMZN & MSFT & GS & WMT & BA & XOM & Mean \\
\midrule
 & Buy-and-hold & −48.3 & −19.3 & −31.2 & −20.6 & −31.1 & −21.9 & −26.6 & −16.1 & −26.9 \\
 & FinMem, 2-y & −44.3 & −21.5 & −25.5 & −14.0 & −31.0 & −14.8 & 0.0 & −21.5 & −21.6 \\
 & FinMem, zero & −40.2 & −18.3 & −17.8 & −22.7 & −31.0 & −13.8 & −30.8 & −16.1 & −23.8 \\
 & FinAgent, 2-y & −30.6 & −20.0 & −34.8 & −13.3 & −20.0 & −19.8 & −8.6 & −23.1 & −21.3 \\
 & FinAgent, zero & −39.1 & −14.5 & −32.7 & −17.7 & −22.2 & −14.6 & −9.0 & −15.6 & −20.7 \\
 & TradingAgents & −27.5 & −11.2 & −13.6 & −18.6 & −16.3 & −15.7 & −10.9 & −8.6 & −15.3 \\
\rowcolor{gray!15} & FinPanel, 2-y & −19.4 & −18.0 & −29.0 & −18.0 & −29.6 & −20.0 & −3.1 & −8.6 & −18.2 \\
\rowcolor{gray!15} & FinPanel, zero & −4.0 & −18.3 & −29.9 & −18.2 & −31.1 & −22.1 & −4.6 & −9.1 & −17.2 \\
\rowcolor{gray!15}\multirow{-9}{*}{2025-H1} & FinPanel, no record & −7.3 & −18.1 & −28.3 & −14.0 & −27.3 & −22.2 & −3.6 & −10.0 & −16.3 \\
\midrule
 & Buy-and-hold & −48.3 & −30.9 & −31.2 & −20.6 & −31.1 & −21.9 & −26.6 & −16.1 & −28.3 \\
 & FinMem, 2-y & −44.3 & −29.5 & −25.5 & −14.0 & −31.0 & −14.8 & −5.2 & −21.5 & −23.2 \\
 & FinMem, zero & −40.2 & −30.3 & −17.8 & −22.7 & −31.0 & −13.8 & −30.8 & −16.1 & −25.3 \\
 & FinAgent, 2-y & −40.0 & −30.0 & −39.0 & −13.3 & −20.0 & −21.4 & −13.7 & −23.1 & −25.1 \\
 & FinAgent, zero & −49.9 & −33.4 & −32.7 & −17.7 & −22.2 & −14.6 & −18.0 & −15.6 & −25.5 \\
 & TradingAgents & −27.5 & −27.0 & −13.6 & −18.6 & −16.3 & −15.7 & −10.9 & −8.6 & −17.3 \\
\rowcolor{gray!15} & FinPanel, 2-y & −19.4 & −25.7 & −29.0 & −18.0 & −29.6 & −20.0 & −15.4 & −8.6 & −20.7 \\
\rowcolor{gray!15} & FinPanel, zero & −5.8 & −24.1 & −29.9 & −18.2 & −31.1 & −22.1 & −15.2 & −9.1 & −19.5 \\
\rowcolor{gray!15}\multirow{-9}{*}{2025-FY} & FinPanel, no record & −12.9 & −26.6 & −28.3 & −14.0 & −27.3 & −22.2 & −20.0 & −10.0 & −20.1 \\
\bottomrule
\end{tabular}

  \caption{Maximum drawdown (\%) per stock and its mean over the eight stocks, on 2025-H1 and 2025-FY; the higher value is the shallower drawdown.}
  \label{tab:mdd}
\end{table*}

\begin{table*}[t]
  \centering\footnotesize\setlength{\tabcolsep}{4pt}
  \begin{tabular}{llrrrrrrrrr}
\toprule
 & & \multicolumn{4}{c}{FinMem/FinAgent stocks} & \multicolumn{4}{c}{Added stocks} & \\
\cmidrule(lr){3-6}\cmidrule(lr){7-10}
Window & Method & TSLA & NFLX & AMZN & MSFT & GS & WMT & BA & XOM & Mean \\
\midrule
 & Buy-and-hold & 100 & 100 & 100 & 100 & 100 & 100 & 100 & 100 & 100 \\
 & FinMem, 2-y & 61 & 80 & 57 & 47 & 98 & 84 & 0 & 84 & 64 \\
 & FinMem, zero & 80 & 84 & 45 & 89 & 98 & 64 & 88 & 81 & 79 \\
 & FinAgent, 2-y & 35 & 69 & 51 & 57 & 65 & 56 & 50 & 59 & 55 \\
 & FinAgent, zero & 33 & 68 & 48 & 54 & 65 & 63 & 50 & 66 & 56 \\
 & TradingAgents & 23 & 20 & 30 & 33 & 25 & 23 & 21 & 22 & 25 \\
\rowcolor{gray!15} & FinPanel, 2-y & 22 & 82 & 86 & 68 & 73 & 83 & 11 & 59 & 60 \\
\rowcolor{gray!15} & FinPanel, zero & 10 & 84 & 71 & 70 & 66 & 84 & 20 & 71 & 60 \\
\rowcolor{gray!15}\multirow{-9}{*}{2025-H1} & FinPanel, no record & 12 & 78 & 86 & 65 & 73 & 72 & 12 & 62 & 58 \\
\midrule
 & Buy-and-hold & 100 & 100 & 100 & 100 & 100 & 100 & 100 & 100 & 100 \\
 & FinMem, 2-y & 72 & 71 & 78 & 47 & 98 & 92 & 4 & 91 & 69 \\
 & FinMem, zero & 90 & 85 & 53 & 92 & 99 & 78 & 94 & 64 & 82 \\
 & FinAgent, 2-y & 37 & 59 & 51 & 68 & 67 & 60 & 52 & 58 & 57 \\
 & FinAgent, zero & 38 & 55 & 41 & 60 & 65 & 70 & 55 & 68 & 57 \\
 & TradingAgents & 23 & 21 & 27 & 33 & 25 & 22 & 25 & 24 & 25 \\
\rowcolor{gray!15} & FinPanel, 2-y & 20 & 76 & 92 & 78 & 75 & 84 & 31 & 67 & 65 \\
\rowcolor{gray!15} & FinPanel, zero & 9 & 80 & 77 & 80 & 72 & 80 & 38 & 73 & 64 \\
\rowcolor{gray!15}\multirow{-9}{*}{2025-FY} & FinPanel, no record & 13 & 78 & 90 & 75 & 73 & 76 & 31 & 69 & 63 \\
\bottomrule
\end{tabular}

  \caption{Exposure per stock, the share (\%) of scored days on which the book holds the stock, and its mean over the eight stocks, on 2025-H1 and 2025-FY.}
  \label{tab:exposure}
\end{table*}

\subsection{Each Trader Alone}
\label{app:solo}
Table~\ref{tab:solo} replays each trader of the default panel as if it were the whole panel, next to the manager's actions and the majority vote of the same calls, with ties between BUY and SELL resolved to HOLD.

\begin{table*}[t]
  \centering\footnotesize\setlength{\tabcolsep}{4pt}
  \begin{tabular}{llrrlrrr}
\toprule
Stock & Window & Manager & Majority & Best seat & best & median & worst \\
\midrule
TSLA & 2025-H1 & 5.0 & −11.6 & Buffett & 4.2 & −14.4 & −26.5 \\
TSLA & 2025-FY & 14.2 & −9.2 & Buffett & 4.2 & −8.1 & −28.0 \\
NFLX & 2025-H1 & 54.6 & 57.0 & Livermore & 59.4 & 52.6 & 37.5 \\
NFLX & 2025-FY & 16.0 & 28.0 & Livermore & 32.8 & 16.8 & −0.8 \\
AMZN & 2025-H1 & −1.9 & −8.5 & Dreman & 3.2 & −3.6 & −7.3 \\
AMZN & 2025-FY & 5.5 & −2.7 & Dreman & 29.4 & 4.6 & −1.4 \\
MSFT & 2025-H1 & 10.0 & 7.5 & Lynch & 17.9 & 8.4 & −3.8 \\
MSFT & 2025-FY & 4.5 & 6.7 & Lynch & 15.8 & 5.8 & −5.6 \\
GS & 2025-H1 & 4.8 & 0.9 & Jones & 9.7 & 0.8 & −6.4 \\
GS & 2025-FY & 35.3 & 37.2 & Jones & 38.9 & 29.9 & 21.8 \\
WMT & 2025-H1 & 8.1 & 8.2 & Shaw & 12.7 & 10.1 & −8.5 \\
WMT & 2025-FY & 24.3 & 19.7 & Simons & 26.9 & 18.5 & 3.1 \\
BA & 2025-H1 & 10.7 & 3.8 & Soros & 23.2 & 8.1 & 0.0 \\
BA & 2025-FY & 12.4 & −4.2 & Dennis & 20.8 & 15.1 & 0.0 \\
XOM & 2025-H1 & −0.6 & 4.8 & Soros & 12.0 & 6.1 & −22.3 \\
XOM & 2025-FY & 12.1 & 18.9 & Soros & 32.1 & 19.8 & −16.0 \\
\bottomrule
\end{tabular}

  \caption{Each trader of the default $N{=}6$ panel replayed alone, cumulative return (\%) without costs, open to open, on 2025-H1 and 2025-FY: the manager's actions, the majority vote of the panel's calls, and each trader as if it were the whole panel (BUY = long from the next open, SELL = flat, HOLD = unchanged); best, median and worst are over the six traders.}
  \label{tab:solo}
\end{table*}

\section{Recruited Panels and the Fixed Panel}
\label{app:roster}
\paragraph{The sketch and the panel for TSLA.} For TSLA over 2023--2024 the sketch reads: annualised volatility 58\%, an average absolute daily move of 2.7\%, 34\% of days moving 3\% or more, an average intraday range of 4.2\% and overnight gap of 1.5\%, and a one-day autocorrelation of $+$0.03. The six names the recruiter seats on it are those shown in Figure~\ref{fig:framework}; the recruited panel of every stock is listed in Table~\ref{tab:panels}.

\paragraph{Ballots.} The names trade differently. Over the 2025-H1 cells of the recruited $N{=}6$ panels, the six seats disagree on 77\% of the scored days and cast all three of BUY, SELL and HOLD on 30\%; the share of BUY calls per seat ranges from 0\% to 93\% with a median of 61\%, 12 of the 48 seats saying BUY on fewer than 30\% of days and 1 on more than 90\%. The fixed panel, whose names were not chosen for the stock, disagrees more often, on 83\% of days, and casts all three calls on 35\%.

\paragraph{The fixed panel.} The w/o selection configuration seats the six names the recruiter seated most often across its recruited panels: Jim Simons, Paul Tudor Jones, George Soros, Richard Dennis, Jesse Livermore and Warren Buffett, in that order on every stock. The recruited panel of each stock shares 4.0 of its six seats with this panel on average (per stock in Table~\ref{tab:panels}).

\begin{table*}[t]\centering\footnotesize\setlength{\tabcolsep}{4pt}\begin{tabular}{lp{0.70\linewidth}r}
\toprule
Stock & Recruited panel ($N{=}6$) & Fixed \\
\midrule
TSLA & Paul Tudor Jones, Jesse Livermore, Jim Simons, George Soros, Nassim Nicholas Taleb, Warren Buffett & 5 \\
NFLX & Ed Seykota, Paul Tudor Jones, Jesse Livermore, Jim Simons, William O'Neil, Richard Dennis & 4 \\
AMZN & Paul Tudor Jones, David Dreman, Warren Buffett, Jim Simons, George Soros, Jesse Livermore & 5 \\
MSFT & Paul Tudor Jones, Linda Bradford Raschke, Jim Simons, Peter Lynch, George Soros, Nassim Nicholas Taleb & 3 \\
GS & Jesse Livermore, Paul Tudor Jones, Ed Thorp, Richard Dennis, George Soros, Jim Simons & 5 \\
WMT & Warren Buffett, Jim Simons, John Paulson, Richard Dennis, David Shaw, Benjamin Graham & 3 \\
BA & Richard Dennis, Jim Simons, Nassim Nicholas Taleb, Paul Tudor Jones, George Soros, Linda Bradford Raschke & 4 \\
XOM & Paul Tudor Jones, Linda Bradford Raschke, Tom DeMark, Ed Seykota, George Soros, Jim Simons & 3 \\
\bottomrule
\end{tabular}
\caption{Recruited $N{=}6$ panels per stock, and the number of seats shared with the fixed panel.}\label{tab:panels}\end{table*}

\clearpage\onecolumn
\section{Prompts}
\label{app:prompts}
The four FinPanel prompts, verbatim; the placeholders in braces are filled with data.
\begin{promptbox}{Recruiter}{blue}
\textbf{System:}\\
You staff panels of traders. Every seat is the same job, worked alone: \\
one look per day at the instrument's daily brief — price and volume, \\
technical indicators, news, filings — and one call: BUY, SELL or HOLD.
\smallskip

\textbf{User:}\\
Staff a panel of \{size\} traders for \{ticker\}.

\{instrument\_block\} \\
Hire \{size\} real, well-known traders or investors, from any era, varied \\
in how they trade and each one suited to this name. For each seat, state \\
first what about the way \{ticker\} trades the hire is suited to, then who \\
it is — one pair of tags per seat, \{size\} pairs in all:

<fit>what about the way \{ticker\} trades this seat is suited to</fit> \\
<name>first and last name</name>
\end{promptbox}

\begin{promptbox}{Trader}{teal}
\textbf{System:}\\
You ARE \{actor\_name\}. Speak in the first person, in your own voice and \\
discipline, making one trading call today. Trade it in your own style and \\
rhythm, as you would have traded the markets of your own era.
\smallskip

\textbf{User:}\\
It is \{date\}. Today's brief for \{ticker\}:

\#\#\#\# Price / factor metrics \\
\texttt{`}\texttt{`}\texttt{`}json \\
\{metrics\_data\} \\
\texttt{`}\texttt{`}\texttt{`}

\{technical\_section\}

\{news\_section\}

\{fundamentals\_section\}

-{}-{}-

What is your call on \{ticker\} today, \{actor\_name\}? Use only what is in the \\
brief. Reasoning first, then your action, each in its own tag:

<reasoning>one paragraph, in your own voice</reasoning> \\
<action>BUY, SELL, or HOLD</action>

BUY = long \{ticker\} now; SELL = flat; HOLD = no view today.
\end{promptbox}

\begin{promptbox}{Distiller}{orange}
\textbf{System:}\\
You characterise an analyst's record: what it keys on, and when its calls \\
tracked the move versus missed it. The record you are given is everything \\
you know about the analyst. Keep the record's numbers in your claims — no \\
vague verdicts. What the record doesn't cover is untested — say so, don't \\
guess. You assess; you don't advise.
\smallskip

\textbf{User:}\\
Below is the analyst's record on \{ticker\} for \{scope\_label\}, newest first — \\
what it argued, and how each call turned out.

\{source\_texts\}

Write ONE paragraph on its judgment of \{ticker\}. Respond with the section \\
wrapped in a single tag:

<section> \\
the characterisation \\
</section>
\end{promptbox}

\begin{promptbox}{Manager}{olive}
\textbf{System:}\\
You are the decision manager of a trading panel: read the day's evidence \\
and commit to ONE final action (BUY, SELL, or HOLD). Do not decide by \\
counting votes: judge what each analyst argued, together with everything \\
else shown, and introduce no outside facts or figures. If the case is \\
bullish, prefer BUY over HOLD; if bearish, prefer SELL over HOLD. Reverse \\
your standing stance only on a changed fact, and name that fact in your \\
reasoning.
\smallskip

\textbf{User:}\\
Decide \{ticker\} today.

What your analysts said today:\{ctx\} \\
\{analyses\_text\} \\
\{records\_section\} \\
Account state (structural facts): \\
\{account\_state\_block\}

Your own recent decisions — the thread you've been committed to: \\
\{memory\_block\}

Reasoning first, then your final action, each in its own tag:

<reasoning>one paragraph: the inputs that drove the call</reasoning> \\
<action>BUY, SELL, or HOLD</action>
\end{promptbox}

\end{document}